\documentclass[11pt,preprint,english,A4]{iopart}

\usepackage{graphicx}
\usepackage{amsmath}
\usepackage{amssymb}
\usepackage{color}
\usepackage{mathtools}
\usepackage{ulem}

\newcommand\msun{\, {M}_\odot}

\newcommand\mseed{\mathcal{M}}

\begin{document}
\title[IMBHs born via repeated mergers are typically thermodynamically unstable]{Intermediate-mass black holes born via repeated mergers are unlikely thermodynamically stable}

\author[0000-0002-7330-027X]{Giacomo Fragione}
\address{Center for Interdisciplinary Exploration \& Research in Astrophysics (CIERA) and Department of Physics \& Astronomy, Northwestern University, Evanston, IL 60208, USA}

\ead{giacomo.fragione@northwestern.edu}

\begin{abstract}
Black hole (BH) thermodynamics is one of the most fascinating aspects of BH physics. While its direct proof is challenging to observe for astrophysical BHs, gravitational waves (GWs) have recently offered a new indirect powerful method to probe that it is operating in the Universe. In this paper, we discuss the thermodynamic stability of IMBHs that are born through repeated mergers of stellar-mass BHs via GW emission in dense star clusters. We show that $\lesssim 20\%$ of the IMBHs born via repeated mergers are thermodynamically unstable for a non-charged BH, while $\lesssim 40\%$ are thermodynamically unstable for a moderately-charged BH. Only in the case of unlikely high charges, $\gtrsim 60\%$ of the IMBHs are thermodynamically stable. With the present and upcoming GW missions, such as LIGO/Virgo/KAGRA, LISA, DECIGO, and ET that promise to detect tens, or even hundreds, of IMBH mergers via GW emission in the next decade, IMBH thermodynamics may soon be tested.
\end{abstract}

\section{Introduction}
\label{sect:intro}

The discovery that black holes (BHs) radiate as black bodies has revolutionized our understanding of general relativity and has offered us some hints about the nature of quantum gravity \cite{Bekenstein1973,Hawking1974,Hawking1975}. BH thermodynamics can be geometrically defined starting with the identifications that entropy and temperature are proportional to the area of the event horizon and surface gravity, respectively. If as in ordinary systems the thermal properties reflect the statistical mechanics of microstates, it may be possible that BH temperature and entropy uncover something about the underlying quantum gravitational states. 

In general, BH thermodynamics is completely defined by the mass, $M$, spin, $J$, and electric charge, $Q$, of the BH. As a result of pair production occurring just outside the event horizon, a BH slowly loses mass as particles are radiated away \cite{Bekenstein1973,Hawking1974,Hawking1975}. It is well established that an isolated Schwarzschild BH is thermodynamically unstable, as a result of its heat capacity being always negative. If thermodynamics plays a role in generating a BH in a stable configuration, then it is likely that a Schwarzschild BH is not the final equilibrium state. Indeed, spinning and/or electrically charged BHs may be thermodynamically stable under some circumstances.  

Although BH thermodynamics is now extensively accepted, the direct observational evidence that it is operating is difficult to be obtained. Indeed, BHs are predicted to emit radiation at the Hawking temperature, which is extremely weak in the case of astrophysical BHs, $T_{\rm H}\lesssim\,10^{-2}\mu$K, far smaller than the temperature of the cosmic microwave background. However, the recent advent of gravitational wave (GW) astronomy has provided a new tool to explore the properties of BHs in strong-field regimes, including BH thermodynamics, e.g.~\cite{CarulloLaghi2021,HuJani2021,IsiFarr2021,WuWei2022}.

Based on their mass and independently on spin and charge, the BH family is typically divided into three subfamilies: stellar-mass BHs ($\lesssim 10^2\msun$), intermediate-mass BHs (IMBHs; $\sim 10^2\msun - 10^5\msun$), and supermassive BHs ($\gtrsim 10^5\msun$). While IMBHs are recognized to have a potential role in cosmology and high-energy astrophysics \cite{GreeneStrader2020}, their origin is still a mystery, owing to the fact that the stars we observe in the Universe collapse to form a BH of maximum $\sim 100\msun$ \cite{bel16b}. While there are a few different venues to produce an IMBH, an efficient way to catalyze its formation is through repeated mergers of stellar-mass BHs via GW emission in dense star clusters, e.g.~\cite{mil02b,antoras2016,FragioneKocsis2021,GonzalezKremer2021,MapelliDall'Amico2021,WeatherfordFragione2021}. 

Current and upcoming GW missions could shed light not only on the origin, but also on the thermodynamics of IMBHs. In this paper, we discuss the thermodynamic stability of IMBHs that are born through repeated mergers of stellar-mass BHs via GW emission in dense star clusters. We compute the fraction of systems that are thermodynamically stable, as a function of the host cluster metallicity and BH spin and electric charge, and show that IMBHs born via repeated mergers are unlikely thermodynamically stable. 

This paper is organized as follows. In Section~\ref{sect:thermo}, we report the main BH thermodynamic quantities we use and discuss the conditions for thermodynamic stability. In Section~\ref{sect:imbh}, we present our results on the thermodynamic stability of IMBHs born via repeated mergers of stellar-mass BHs in dense star clusters. Finally, in Section~\ref{sect:concl}, we discuss the implications of our results and draw our conclusions.

\section{Black Hole Thermodynamics}
\label{sect:thermo}

In this Section, we summarize the most relevant BH thermodynamic quantities for BHs we use and discuss the conditions for thermodynamic stability.

The relation between the geometrically-defined BH thermodynamic quantities and their classical counterparts is still unknown. Nevertheless, it is possible to apply traditional thermodynamic concepts to Kerr-Newman BHs \cite{Davies1977,Carlip2014}. In what follows, we use Planck units, with the exception of setting the value of the Boltzmann constant to $1/8\pi$. The fundamental thermodynamic equation is \cite{Smarr1973}
\begin{equation}
    M(S,J,Q)=\left(2S+\left(1/8S\right)\left(J^2+Q^4/4\right)+Q^2/2\right)^{1/2}\,,
\end{equation}
where $S$ is the BH entropy. In analogy with ordinary thermodynamics, the first law of BH thermodynamics states that if $M$ changes by an infinitesimal quantity $dM$, then
\begin{equation}
    dM=TdS+\Omega dJ+\Phi dQ\,,
\end{equation}
where the temperature $T$, angular momentum $\Omega$, and electric potential $\Phi$ are defined as
\begin{eqnarray}
    T&=&\left.\left(\frac{\partial M}{\partial S}\right)\right|_{J,Q}=\frac{1}{M}\left[1-\frac{J^2+Q^4/4}{16S^2}\right] \\
    \Omega&=&\left.\left(\frac{\partial M}{\partial J}\right)\right|_{T,Q}=\frac{J}{8MS} \\
    \Phi&=&\left.\left(\frac{\partial M}{\partial Q}\right)\right|_{T,J}=\frac{Q\left(Q^2+8S\right)}{16MS}\,.
\end{eqnarray}

\begin{figure} 
\centering
\includegraphics[scale=0.75]{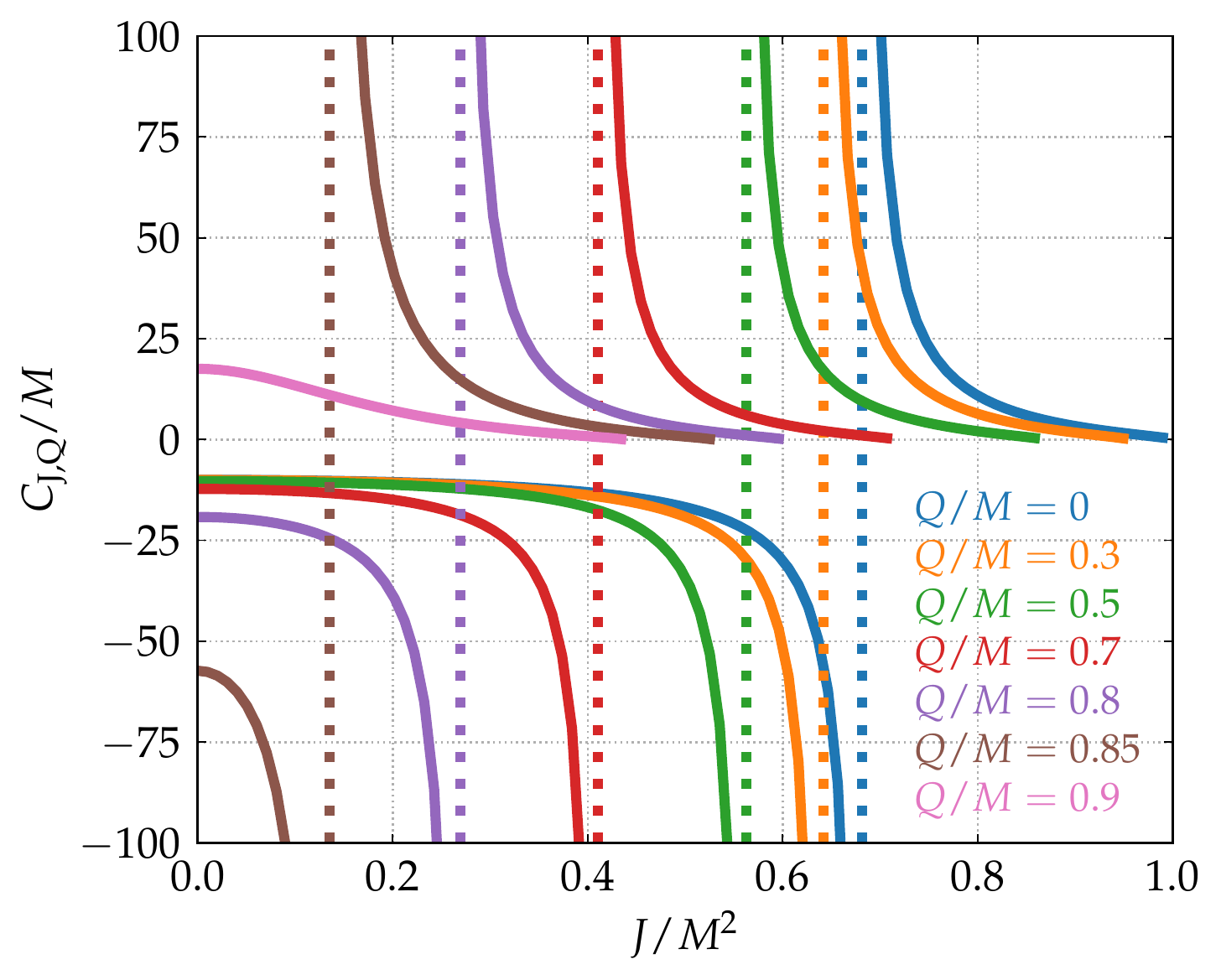}
\caption{Solid lines: BH thermal capacity for different spins and electric charges. Dotted lines: discontinuity at which the BH heat capacity changes from positive to negative for different spins and electric charges.}
\label{fig:cjq}
\end{figure}

As in classical thermodynamics, the various thermal capacities have a crucial importance. If a BH is in equilibrium at some temperature $T$ in a thermal bath, the BH will absorb energy reversibly if the temperature of the external bath is increased by an amount $dT$. The absorption will be isotropic, therefore the angular momentum and electric charge will not change on average. In this case, we can define the thermal capacity at constant $J$ and $Q$ as
\begin{equation}
    C_{J,Q} = T\left.\left(\frac{\partial S}{\partial T}\right)\right|_{J,Q}=\frac{8MS^3T}{J^2+Q^4/4-8T^2S^3}\,.
    \label{eqn:specheat}
\end{equation}
Eq.~\ref{eqn:specheat} has a singularity whenever the denominator vanishes, marking a phase transition. Using $J^2=\alpha M^4$ and $Q^2=\beta M^2$, this happens whenever \cite{Davies1977}
\begin{equation}
    \alpha^2+6\alpha+4\beta-3=0\,.
\end{equation}
If the equation above is defined positive, then $C_{J,Q}>0$, otherwise $C_{J,Q}<0$ and the system becomes hotter as it radiates energy. Figure~\ref{fig:cjq} shows the BH heat capacity for different spins and electric charges. We also report the location (dotted lines) of the discontinuity at which the BH heat capacity changes from positive to negative, marking a phase transition, for different spins and electric charges. For a Schwarzshild BH, $C_{J,Q}\propto -M^{2}$. Typically, for high values of the BH spin, $C_{J,Q}$ is positive, while for low values of the BH spin, $C_{J,Q}$ is negative. The heat capacity is always positive for larger BH electric charges, that is $Q/M\gtrsim 0.9$. Indeed the value of $Q$ affects the location of the discontinuity: the smaller the electric charge is, the larger is the spin at which the phase transition occurs.

As in ordinary systems, the condition for equilibrium is that the Helmotz free energy,
\begin{equation}
    F=M-TS\,,
\end{equation}
has a minimum. This translates into 
\begin{equation}
    (4\alpha+\beta^2)(\alpha^2+6\alpha+4\beta-3)>0\,,
\end{equation}
which implies $C_{J,Q}>0$ for thermodynamic stability. Therefore, a BH is in a stable thermodynamic equilibrium whenever its thermal capacity is defined positive. The stability or instability will finally affect the evaporation process and the time evolution of the BH entropy and mass, e.g.~\cite{Page2013,Aref'evaVolovich2022}.

\section{Thermodynamic stability of intermediate-mass black holes}
\label{sect:imbh}

In this Section, we discuss the thermodynamic stability of IMBHs born as a result of repeated mergers of stellar-mass BHs.

In what follows, we use the results obtained in \cite{FragioneKocsis2021} for the formation and growth of IMBHs, starting from a BH seed of mass $\mseed$. Before discussing our results, we briefly summarize the overall dynamical processes that shape the origin of an IMBH through repeated mergers in a dense stellar environment. The initial BH seed (that will grow to form an IMBH) is born as a result of the stellar evolution of its progenitor massive star. The latter can either be one of the stars that formed in the beginning of the cluster lifetime or a star originated as a result of mergers of smaller stars in the cluster center (which depends on the initial cluster density and mass, e.g.~\cite{GonzalezKremer2021,WeatherfordFragione2021}). Given a stellar progenitor, the mass of the BH seed (and of the other BHs in the cluster) is essentially set by the cluster metallicity, e.g.~\cite{HurleyPols2000,BanerjeeBelczynski2020}. If the seed is not ejected by natal kicks (as a result of anisotropy when the star collapses in a supernova event), it sinks to the cluster center via dynamical friction \cite{Chandrasekhar1943}. Owing to the large densities in the cluster core, the BH seed soon forms a binary with another BH ($m_2$) via three-body or binary-single encounters, whichever is the faster depending on the cluster mass and density, e.g.~\cite{antoras2016}. After the binary is formed, it shrinks as a result of encounters with other BHs and stars \cite{quin1996}, to the point where eventually GWs take over and the system merges via GW emission \cite{peters64}. If not ejected from the cluster as a result of relativistic recoil kicks from asymmetries in GW emission, e.g.~\cite{lou12,HofmannBarausse2016,Jimenez-FortezaKeitel2017}, the merger product is retained within the cluster, can form a new binary, and eventually keep merging to form an IMBH.

\begin{figure} 
\centering
\includegraphics[scale=0.625]{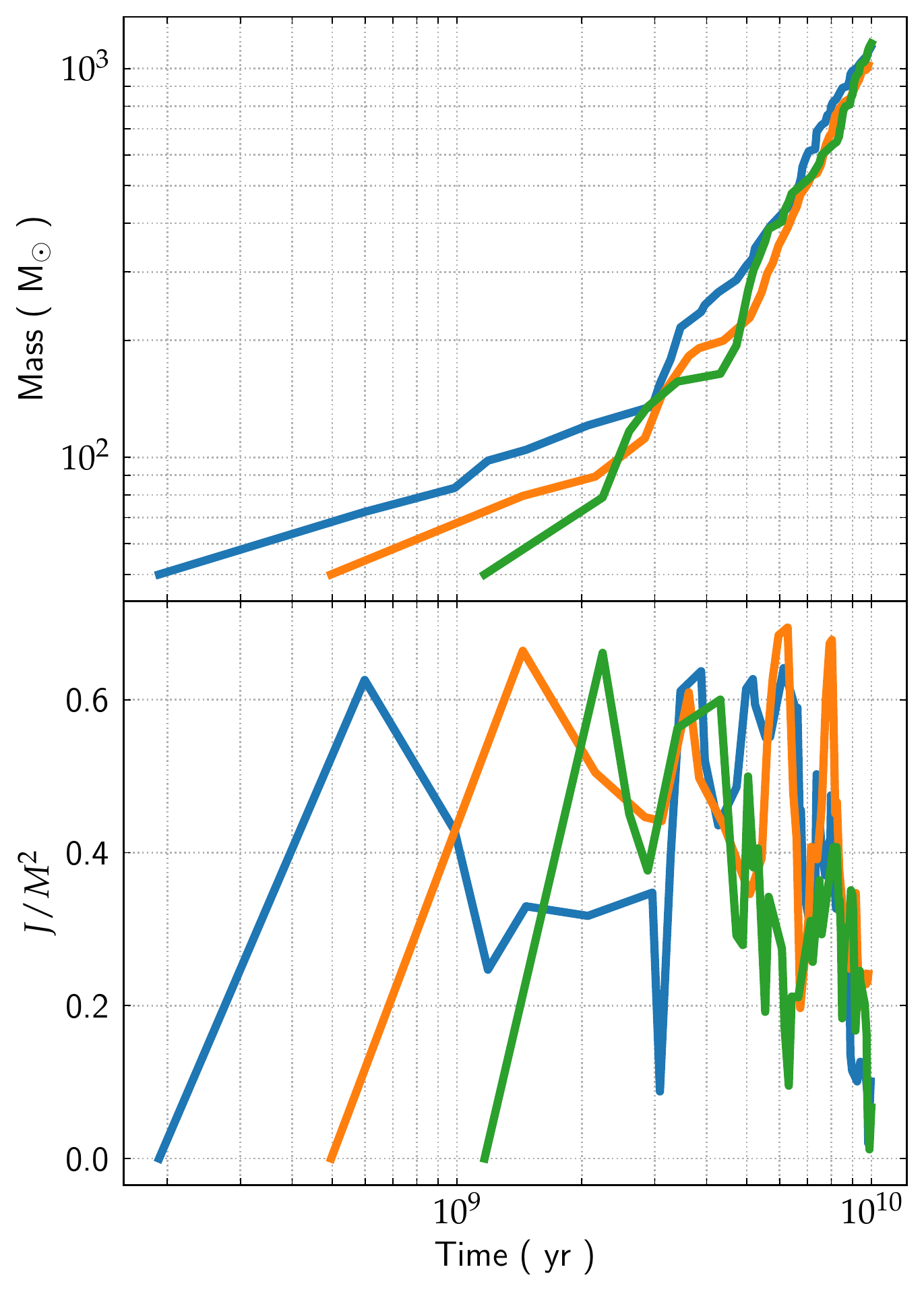}
\caption{Example of growth and spin evolution of $\sim 1000\msun$ IMBH from the runs of \cite{FragioneKocsis2021}, for star clusters with metallicity $Z=0.0001$. Different colors represent different formation histories. The initial BH seed mass is $50\msun$ and the initial BH spins are assumed to be zero.}
\label{fig:example}
\end{figure}

\begin{figure} 
\centering
\includegraphics[scale=0.625]{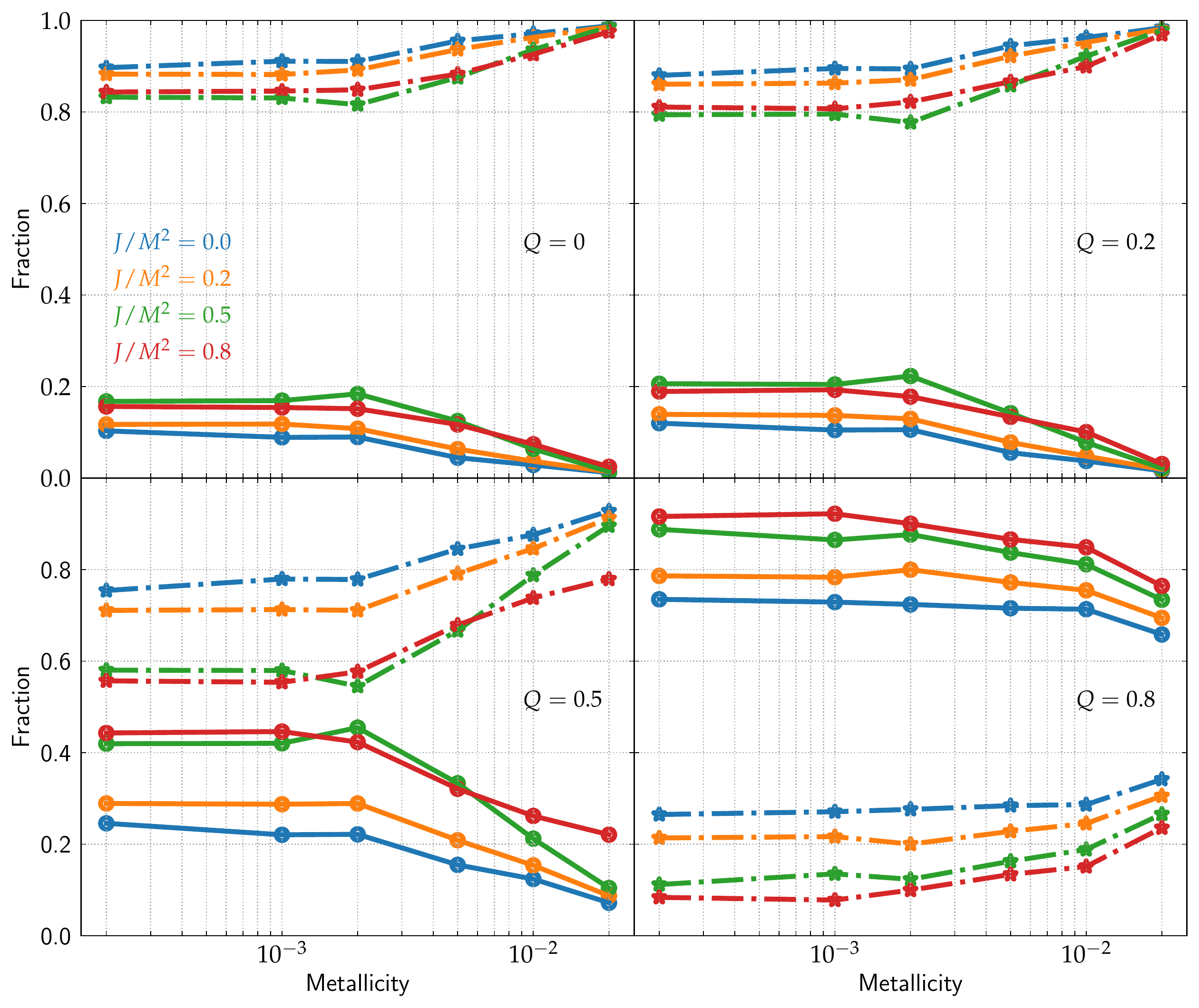}
\caption{Fraction of IMBHs born via repeated mergers over cosmic time that are thermodynamically stable (solid lines) and unstable (dashed lines) as a function of the cluster metallicity. The initial BH seed mass is $50\msun$. Different colors represent different initial BH spins. Different panels represent different values of the electric charge: $Q=0$ (top-left panel), $Q=0.2$ (top-right panel), $Q=0.5$ (bottom-left panel), $Q=0.8$ (bottom-right panel).}
\label{fig:stability}
\end{figure}

We show an example of growth of $\sim 1000\msun$ IMBHs in Figure~\ref{fig:example}. We illustrate the mass evolution and spin evolution as a function of time, for star clusters with metallicity $Z=0.0001$. The initial BH seed mass is $50\msun$ and the initial BH spins are assumed to be zero. It is clear that the spin of the growing seed decreases as a function of time (or its growing mass), e.g.~\cite{antonini2019,frsilk2020}. Indeed, after the first merger producing a remnant with a spin parameter ($J/M^2$) of about $0.7$, e.g.~\cite{lou12,HofmannBarausse2016,Jimenez-FortezaKeitel2017}, the remnant spin tends to decrease with subsequent mergers. This trend comes from the fact that the angular momentum imparted to the growing seed during each merger is $\propto m_2\mseed$, which causes the spin parameter to change by $\sim m_2/\mseed$. Assuming an isotropic geometry of binary BH mergers as appropriate to a dynamical environment, the final inspiral and deposition of angular momentum happens at random angles with respect to the spin axis of the growing seed. Moreover, since at early times $m_2$ can be comparable to $\mseed$, the initial variations in the remnant spins tend to be larger than the ones at late times. The growing seed undergoes a damped random walk in the evolution of its spin since retrograde orbits become unstable at a larger specific angular momentum than do prograde orbits. As a consequence, the magnitude of the spin parameter tends to decrease \cite{Miller2002,HughesBlandford2003,MandelBrown2008}.

Figure~\ref{fig:stability} shows the fraction of IMBHs born via repeated mergers (over cosmic time) that are thermodynamically stable (solid lines) and unstable (dashed lines) as a function of the metallicity of the host cluster. In these simulations, we have fixed the initial BH seed mass to $50\msun$, while we have considered different initial BH spins, from a non-spinning case ($J/M^2=0$) to a highly-spinning case ($J/M^2=0.8$). Since the thermodynamic stability of a BH could critically depends on its electric charge, we show in different panels the results for different BH electric charges, from a neutral BH ($Q=0$) to a highly-charged BH ($Q=0.8$). We find that $\lesssim 20\%$ of the IMBHs born via repeated mergers are thermodynamically unstable for $Q<0.2$, while $\lesssim 40\%$ are thermodynamically unstable for $Q<0.5$. Only in the case $Q=0.8$, $\gtrsim 60\%$ of the IMBHs are stable (see Figure~\ref{fig:cjq}). However, the case of a highly-charged BH should be very unlikely in Nature. As expected, we also find that the fraction of thermodynamically-stable IMBHs is larger for initial larger BH spins, $J/M^2>0.5$, typically twice the fraction of the case of smaller initial BH spins. Finally, there is also a trend with the metallicity of the host cluster, with the fraction of stable systems being smaller at larger metallicities. The reason is that smaller metallicities lead to more massive BHs ($m_2$), which can affect the spin parameter of the growing seed more importantly, being its change at every merger proportional to $m_2$.

\section{Discussion and Conclusions}
\label{sect:concl}

The surprising discovery that BHs behave as thermodynamic objects has radically affected our understanding of general relativity and its relationship to quantum field theory. Since the Hawking temperature is extremely weak in the case of astrophysical BHs, the direct observation that BH thermodynamics is operating in nature is quite challenging. However, the recent advent of GW astronomy has provided a new tool to explore the properties of BHs in the strong-field regimes, including BH thermodynamics.

In this paper, we have discussed the thermodynamic stability of IMBHs that are born through repeated mergers of stellar-mass BHs via GW emission in dense star clusters. We have shown that $\lesssim 20\%$ of the IMBHs born via repeated mergers are thermodynamically unstable for $Q<0.2$, while $\lesssim 40\%$ are thermodynamically unstable for $Q<0.5$. Only in the unlikely case $Q=0.8$, $\gtrsim 60\%$ of the IMBHs are stable.

The implications of the thermodynamic instability of astrophysical BHs are still far from being understood. However, present and upcoming GW missions, such as LIGO/Virgo/KAGRA, LISA, DECIGO, and ET that promise to detect tens, or even hundreds, of BH mergers via GW emission in the next decade, possibly shedding light on BH thermodynamics.

\ack{We thank Avi Loeb and Fred Rasio for useful discussions. G.F.\ acknowledges support from NASA Grant 80NSSC21K1722.}

\section*{References}
\bibliographystyle{utphys}
\bibliography{refs}

\end{document}